**Telecom AI Native Systems in the Age of Generative AI – An Engineering Perspective**

Authors:
- Ricardo Britto, Ericsson and Blekinge Institute of Technology, Sweden
- Timothy Murphy, Ericsson, Canada
- Massimo Iovene, Ericsson, Italy
- Leif Jonsson, Ericsson, Sweden
- Melike Erol-Kantarci, Ericsson, Canada
- Benedek Kovács, Ericsson, Hungary

# 1. Introduction

In recent years, Artificial Intelligence (AI) has seen astonishing advancements in dealing with vast amounts of data and generating actionable insights. The impact on various industries has been visible and remarkable, influencing, improving, and sometimes revolutionizing entire sectors. In recent months, the rise of generative AI and its ability to generate new content, e.g., natural language text or video, based on a given input or context [1], has exponentially increased the expectation of what AI can deliver in terms of innovation, industry transformation, new business opportunities, and operations simplification.

Many generative AI models are packaged as Foundation models (FM). FMs come in various types. There are foundation models for text, image, sound, and video, but the most well-known FMs are text-oriented models called large language models (LLMs). LLMs are artificial deep neural networks that can generate new text-oriented data. They are trained on massive amounts of text data from various sources, such as online books, news articles, social media posts, programming code, and web pages. LLMs have shown impressive capabilities in various natural language processing (NLP) tasks, such as text summarization, question answering, sentiment analysis, code generation, and machine translation. They can also generate creative and engaging texts, such as stories, poems, jokes, lyrics, images, audio, and code, all driven by text-based prompts.

From an industry perspective, FMs can revolutionize how we interact with software products and services. They can enable new forms of human-computer communication, such as conversational agents, and personal assistants. They can also enhance software products' functionality and user experience, such as search engines, e-commerce platforms, and social media networks. The rise of products such as Open AI's ChatGPT [2] shows how profound the impact of this type of technology can be on society.

The transformational power of AI is evident across many businesses, including the telecom industry. The increasing relevance of AI in telecom use cases has been observed for some years, leading to the rise of the term AI native telco. A recent Ericsson whitepaper [3] explains the term AI native as a system that has *"intrinsic trustworthy AI capabilities, where AI is a natural part of the functionality in design, deployment, operation, and maintenance. An AI native system capitalizes on a data-driven and knowledge-based ecosystem, where data is created and consumed to produce new AI-based functionality, replacing static, rule-based mechanisms with learning and adaptive AI when needed"* [3]. The power and flexibility of FMs make them an obvious building block of AI native systems.

Developing software products incorporating FM components may introduce legal and intellectual property rights (IPR) issues and additional engineering complexity. The stochastic nature of FMs, the data quality, the model size, trustworthiness, security, regulatory, and privacy aspects [2] amplify the challenges associated with a software's life cycle. Lu et al. have called for action to focus on the design aspects of foundation model-based systems [4], but this area requires more attention from both research and practice communities.

This article reflects on AI native systems leveraging FMs in telecommunication networks and associated implications from an engineering perspective.

## 2. AI in Telecom

FMs (especially LLMs) have become prevalent very quickly in the so-called consumer market, also known as business-to-consumer (B2C), where companies have access to their customers directly (and the associated data). This market segment is very different from the business-to-business (B2B) segment, where companies like Ericsson play most of the time. Communication Service Providers (CSPs) usually play in the B2C market, while suppliers like Ericsson have the CSPs as main customers (B2B).

In B2C, things like chat support deliver immediate results to end users, in most cases exceeding the expectation of the requestor. Since non-mission-critical use cases have been the main part of the first wave of B2C LLM-based products, a failure in the AI system (e.g., an incorrect recommendation) may not completely ruin the use case but might only lower the user experience. Furthermore, the companies in B2C usually have direct access to the data associated with their products, enabling quick innovation and validation of new ideas.

In contrast to most products offered in B2C, the products developed by Telecom suppliers, e.g., Radio Access Network (RAN) [5], have stringent system requirements and are less tolerant of an incorrect answer provided by AI models. This aspect is further amplified by the complexity of telecom ecosystems, which include many systems with real-time constraints. Although tasks like making a mobile phone call may look simple from an end user's perspective (e.g., a CSP's subscriber), they require complex interactions between many systems. These interactions can be disrupted if any of those systems are affected by, for instance, unreliable AI models, and any degradation in system performance is unacceptable to the CSP's.

In addition to the inherent complexity of telecom ecosystems, it is often the case that it is not straightforward to obtain the data associated with those systems, either due to the complexity of simulating systems' behavior in a lab environment or legal and privacy aspects connected to getting data from CSPs.

Despite the obstacles mentioned above, AI approaches (such as FMs) are still necessary to address use cases where procedural or purely deterministic approaches do not fit. Thus, it is essential to understand how AI can be introduced in Telecom products, especially the necessary adaptations or additions to software engineering practices.

Recently, Ericsson has presented a view of the evolution of AI native Telecom [3] (see Figure 1), where AI takes center stage; AI capabilities are not relegated to drop-in functional replacements but are intrinsic components of the software's capabilities. Design, deployment, operation, and maintenance must be aligned with a new AI native-based approach, significantly departing from traditional end-to-end engineering flows.

AI native applications leverage a data-driven, knowledge-based ecosystem; information is consumed and dynamically generated to realize new AI-based functionalities or augment and replace static mechanisms. As such, FMs have an unmeasurable potential to enable AI native applications, even more so in the upcoming Sixth Generation of Mobile Networks (6G), where connectivity, capacity, latency, mobility, and reliability requirements will be harder to fulfill without AI [6].

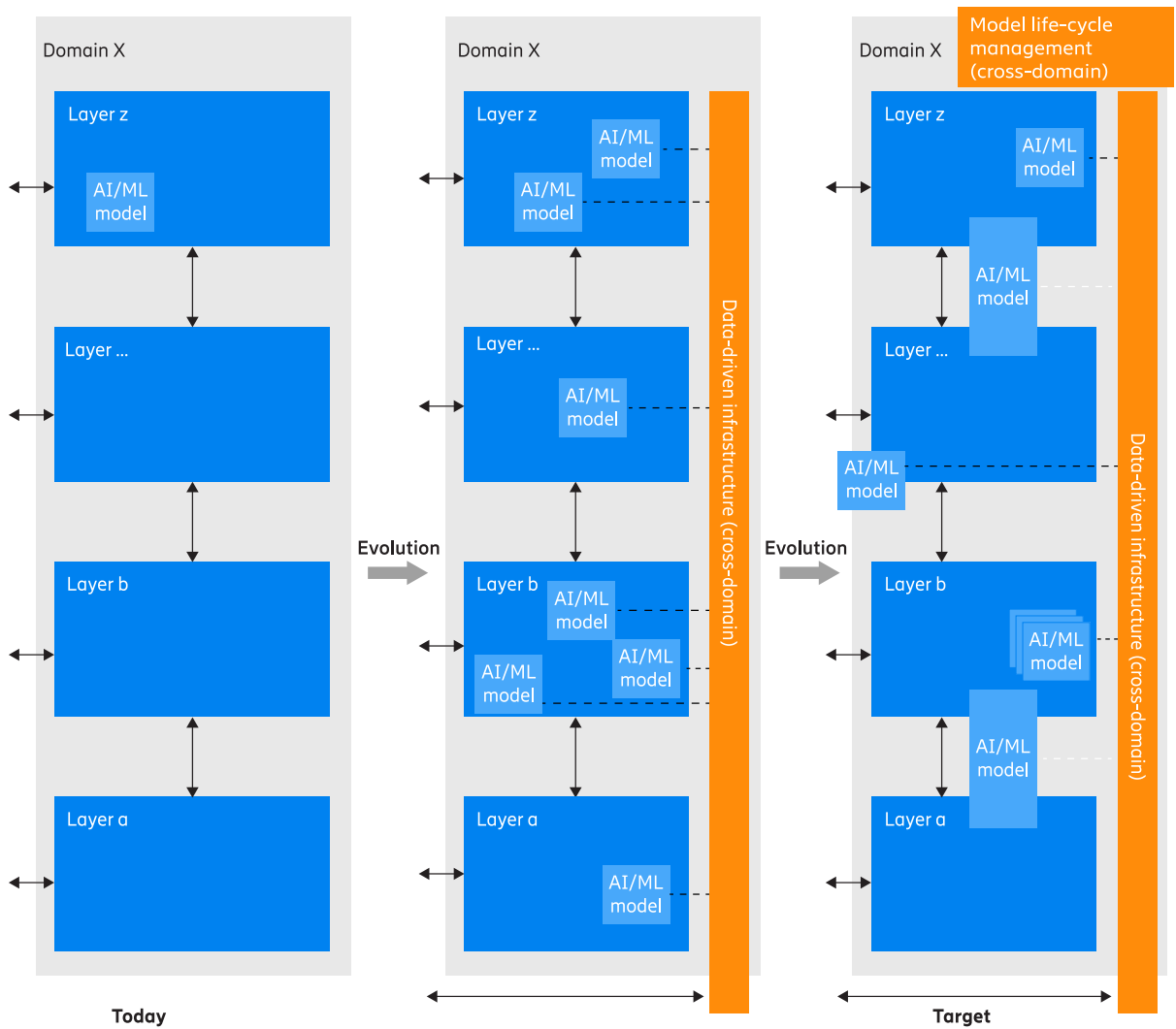

Figure 1: Intelligence everywhere across the architecture

Some examples of use cases where FMs are potentially a good match in mobile networks are as follows [7]:
- **Channel Quality Estimation** – FMs are a good alternative to determine the state of a wireless channel in a mobile network, especially in massive multiple-input multiple-output (MIMO).
- **Network traffic analysis and anomaly detection** – FMs have also been used successfully for detecting and predicting anomalies. They can also be used to recommend corrective actions.

## 3. Engineering Considerations

Developing AI native applications, especially ones including FMs, brings a set of unique considerations for the overall software life cycle. For AI systems in general, the MLOps (Machine Learning Ops) paradigm has existed for some time. It is an extension of DevOps, where development and operations are synergically planned and executed.

In the case of software systems with FMs, the software life cycle has some particularities. Rather than emphasizing training new models from scratch, engineers focus on pre-trained FMs, using different strategies to accomplish their desired goals (e.g., prompt engineering, in-context learning, and fine-tuning). Considering the specific needs of FMs, the community has proposed an extension to MLOps called FM/LLMOps. In this context, we highlight some specific engineering challenges associated with the life cycle of AI native applications, especially the ones using FMs.

In the **development phase**, requirements and design must explicitly cater to AI capabilities. This requires a deep understanding of the problem domain, the data available, and the potential AI models that could be employed. Considering AI during the very beginning (requirements engineering) ensures AI is not simply bolted onto a pre-existing structure. Rather, it is seamlessly integrated into the core architecture of the application.

Activities that demand particular attention in this phase are eliciting, specifying, and verifying requirements, particularly Non-Functional Requirements (NFRs); existing evidence indicates additional complexity associated with requirements engineering for AI systems [7]. To complicate matters further, mature tools and techniques are still lacking to engineer AI systems [8].

It cannot be overlooked that additional integration challenges are associated with incorporating FMs in software products; most often, companies use proprietary models consumed via APIs (e.g., Open AI models) or open-source models deployed independently. In both cases, changes in the underlying models or their exposure might not be backward compatible. This can be problematic since backwards compatibility is typically strictly managed in Telecom systems.

While most AI approaches are non-deterministic by nature, the flexibility associated with FMs amplifies the challenges associated with the validation and verification of AI systems; the considerable number of parameters enable FMs to handle multiple downstream tasks, making it more complicated to secure software quality due to the myriads of behaviors (including "hallucinations") to be verified and validated. That becomes even more challenging in the telecom context due to the mission-critical nature of most software systems in this domain.

Intrinsic to the core of AI native SW engineering flows, development must thoroughly consider foreseeable ethical implications (e.g., algorithms and data to be fair and unbiased). As AI becomes increasingly integral to an application's functionality, ensuring it behaves ethically and responsibly, in a predictable and verifiable manner, becomes essential to verification and validation activities. Additional development costs will likely be related to properly testing and documenting AI operations for transparency and accountability. A fundamental problem here is to verify the enormous space of possible outputs of FM-based systems. How to do this reliably is still an open problem.

The **deployment phase** also contains its share of challenges. While traditional AI approaches require frequent retraining, FMs benefit from their vast number of parameters and leverages techniques such as LoRA (Low-Rank Adaptation of Large Language Models) [9] and RAG (Retrieval-Augmented Generation) [10] to avoid expensive full retraining, a challenge with this approach though, can be to maintain consistent behavior across deployment versions.

Additionally, these models typically require GPUs to achieve good performance, which tends to be more expensive, both in direct HW cost and energy consumption, than "traditional" infrastructure (relevant for the cases where companies decide to deploy FMs independently), so balancing performance, scaling, and cost in deployment is a challenge. It is also worth mentioning that the software ecosystem for deploying FMs is not as mature and flexible as it is for traditional machine learning systems.

In the **operational phase**, AI models require ongoing monitoring to ensure optimal performance and, potentially, optimized model update activities. This involves setting up robust monitoring to utilize "meta-feedback" loops that monitor application performance and performance change. Some additional challenges connected to FM operations are, for instance, performance evaluation and latency. The hold-out approach used with "traditional" machine learning approaches is unsuitable for FMs. Instead, A/B testing is the most used approach. From a telecom perspective, the latency may complicate the use of FMs for mission-critical telecom use cases, even if FMs make sense from a functional perspective. A/B testing can be challenging to execute during development due to the nature of the telecom market and the more prominent business model (B2B).

## Final Thoughts

AI has a pivotal role in the telecom industry, and the possibilities brought by modern foundational models are undeniable. Their power and flexibility enable the realization of many valuable use cases in the telecom domain.

While the potential is out there to be harvested by the industry, engineering challenges complicate introducing this technology, especially in the telecom industry. Formal engineering practices are still missed for some of the key activities of the software life cycle management. Furthermore, operations can be expensive and more complex than operating "traditional" systems.

To unleash the full potential of AI, especially FMs, in telecom, it is essential to take an AI native-first approach and be mindful of all associated challenges, increasing the chances of succeeding in the very competitive telecom market.

## References

[1] R. Bommasani, D. A. Hudson, E. Adeli, R. Altman, S. Arora, S. von Arx, M. S. Bernstein, J. Bohg, A. Bosselut, E. Brunskilletal, "On the opportunities and risks of foundation models," arXiv preprint arXiv:2108.07258, 2021.
[2] E. A. van Dis, J. Bollen, W. Zuidema, R. van Rooij, and C. L. Bockting, "Chatgpt: five priorities for research," Nature, vol. 614, no. 7947, pp. 224–226, 2023.
[3] M. Iovene, L. Jonsson, D. Roeland, M. D'Angelo, G. Hall, M. Erol-Kantarci, J. Manocha, "Defining AI native: A key enabler for advanced intelligent telecom networks," available at: www.ericsson.com/en/reports-and-papers/white-papers/ai-native
[4] L. Q. Lu, Z. Zhu, Z. Xu, Z. Xing, J. Whittle, "A framework for designing foundation model-based systems," arXiv preprint arXiv: 2305.05352, 2023.
[5] A. Damola, Z. Filipovic, "How best to apply AI in the Intelligent RAN Automation, " available at: www.ericsson.com/en/blog/2022/3/applying-ai-in-the-intelligent-ran-automation-for-innovation
[6] H. Tataria, M. Shafi, A. F. Molisch, M. Dohler, H. Sjöland, F. Tufvesson: 6G Wireless Systems: Vision, requirements, challenges, insights, and opportunities. Proceedings of the IEEE 109(7), 1166–1199 (2021)
[7] A. Karapantelakis, P. Alizadeh, A. Alabassi, K. Dey, A. Nikou, "Generative AI in mobile networks: a survey," Annals of Telecommunications (2023).
[8] J. Horkoff: "Non-functional requirements for machine learning: Challenges and new directions," in IEEE 27th International Requirements Engineering Conference (RE), 2019.
[8] Görkem Giray. A software engineering perspective on engineering machine learning systems: State of the art and challenges. Journal of Systems and Software, 180:111031, 2021.
[9] E. J. Hu, Y. Shen, P. Wallis, Z. Allen-Zhu, Y. Li, S. Wang, L. Wang, W. Chen: "LoRA: Low-Rank Adaptation of Large Language Models,"
[10] P. Lewis, E. Perez, A. Piktus, F. Petroni, V. Karpukhin, N. Goyal, H. Küttler, M. Lewis, W. Yih, T. Rocktäschel, S. Riedel, Douwe Kiela "Retrieval-Augmented Generation for Knowledge-Intensive NLP Tasks," in 34th Conference on Neural Information Processing Systems (NeurIPS2020), 2020.